\documentclass[12pt]{article}
\usepackage{array}
\usepackage{graphicx}
\usepackage{amssymb}
\usepackage{amsmath}
\input{epsf}
\usepackage{cite}
\def\@fmsl@sh#1#2#3{\m@th\ooalign{$\hfil#1\mkern#2/\hfil$\crcr$#1#3$}}
 \def\eq#1\en{\begin{equation}#1\end{equation}}
\def\s[#1,#2]{[#1\stackrel{\star}{,}#2]}
\def\sx[#1,#2]{[#1\stackrel{\star_{x}}{,}#2]}

\newcommand{\nc}{\newcommand}
\nc{\beq}{\begin{equation}}
\nc{\eeq}{\end{equation}}
\nc{\beqa}{\begin{eqnarray}}
\nc{\eeqa}{\end{eqnarray}}

\def\bc{\begin{center}}
\def\ec{\end{center}}

\def\to{\rightarrow}

\def\gsim{\mathrel{\mathpalette\atversim>}}

\def\bc{\begin{center}}
\def\ec{\end{center}}

\def\gsim{\mathrel{\rlap{\lower4pt\hbox{\hskip1pt$\sim$}}

    \raise1pt\hbox{$>$}}}       

\def\gsim{\mathrel{\rlap{\lower4pt\hbox{\hskip1pt$\sim$}}
    \raise1pt\hbox{$>$}}}       

\def\simlt{\mathrel{\lower2.5pt\vbox{\lineskip=0pt\baselineskip=0pt
           \hbox{$<$}\hbox{$\sim$}}}}

\begin{document}
\makeatletter
\def\fmslash{\@ifnextchar[{\fmsl@sh}{\fmsl@sh[0mu]}}
\def\fmsl@sh[#1]#2{%
  \mathchoice
    {\@fmsl@sh\displaystyle{#1}{#2}}%
    {\@fmsl@sh\textstyle{#1}{#2}}%
    {\@fmsl@sh\scriptstyle{#1}{#2}}%
    {\@fmsl@sh\scriptscriptstyle{#1}{#2}}}
\def\@fmsl@sh#1#2#3{\m@th\ooalign{$\hfil#1\mkern#2/\hfil$\crcr$#1#3$}}
\makeatother

\thispagestyle{empty}
\begin{titlepage}
\begin{flushright}
{\tt CERN-PH-TH/2011-218}\\
\end{flushright}
\boldmath
\begin{center}
  \Large {\bf Brane World Models Need Low String Scale}
    \end{center}
\unboldmath
\vspace{0.2cm}
\begin{center}
 {\large I. Antoniadis}\footnote{ignatios.antoniadis@cern.ch}\footnote{On leave from CPHT (UMR CNRS 7644) Ecole Polytechnique, F-91128 Palaiseau}$^a$,
{\large M. Atkins}\footnote{m.atkins@sussex.ac.uk}$^b$
 and 
{\large X. Calmet}\footnote{x.calmet@sussex.ac.uk}$^b$
 \end{center}
\begin{center}
{\sl $^a$ Department of Physics, CERN - Theory Division, CH-1211 Geneva 23, Switzerland
\\
$^b$ Physics and Astronomy, 
University of Sussex,  Falmer, Brighton, BN1 9QH, UK 
}
\end{center}
\vspace{2cm}
\begin{abstract}
\noindent
Models with large extra dimensions offer the possibility of the Planck scale being of order the electroweak scale, thus alleviating the gauge hierarchy problem. We show that these models suffer from a breakdown of unitarity at around three quarters of the low effective Planck scale. An obvious candidate to fix the unitarity problem is string theory. We therefore argue that it is necessary for the string scale to appear below the effective Planck scale and that the first signature of such models would be string resonances. We further translate experimental bounds on the string scale into bounds on the effective Planck scale. 
\end{abstract}  

\end{titlepage}



\newpage
The discovery that the scale of quantum gravity could be as low as a few TeV if there are large extra dimensions has been one of the most exciting theoretical developments of the last 15 years \cite{ArkaniHamed:1998rs, Antoniadis:1998ig}. Models with large extra dimensions allow for a geometric reformulation of the gauge hierarchy problem. Typically, in these models, gravity propagates in the full $d$-dimensional space-time while matter fields are confined to a 3-brane. Because the extra dimensions have to be compact, one expects from a four dimensional effective field theory perspective the presence of a tower of massive Kaluza Klein (KK) gravitons. These models offer the possibility of being able to observe quantum gravity, such as string theory, in the very near future at the LHC.

The gauge hierarchy problem is alleviated by proposing that the effective Planck scale, $M_D$, is near the weak scale. Given $n$ extra space dimensions compactified on a torus of common radius $R$, the four dimensional reduced Planck mass is related to the low effective Planck scale via $\bar M_P^2=R^n M_D^{2+n}$. With no UV completion to the theory, it is well known that for $n>1$ extra dimensions these models suffer from UV divergences even at tree level (due to the asymptotic high density of the infinite tower of KK modes). However, one generally expects that quantum gravity at the weak scale will tame these divergences. 

One of the clearest ways to see the divergent behaviour is in the breakdown of tree level unitarity. Scattering amplitudes grow with energy, eventually exceeding bounds derived from the unitarity of the $S$ matrix. However there is a further feature of these models which can also cause problems with unitarity. This feature appears due to the very fine spacing of the KK modes which essentially means that scattering via graviton exchange will involve the exchange of on shell gravitons at whatever energy the scattering takes place. Resonance behaviour due to the transfer of on shell gravitons can push the size of the amplitude beyond the unitarity bound.

In this letter we review the results of \cite{Atkins:2010re} which show that, in general, models with large extra dimensions violate tree level unitarity at around $\frac12 M_D$. Here we investigate the nature of this result further and show that it in fact occurs due  to the resonance behaviour mentioned above and is very sensitive to the cutoff. With this understanding, we show that a far more robust bound can be derived by considering only the imaginary part of the amplitude which encapsulates only the resonant behaviour. From this we find that for $n>1$ extra dimensions, unitarity is violated at around $0.8M_D$. Although this bound is not as low as that found previously, it is much less sensitive to the cutoff and we therefore consider it to be on a much stronger footing.  From this we argue that if string theory is the theory of quantum gravity, we  would require the string scale, $M_S$, to be at or below the scale of unitarity violation. The breakdown of unitarity therefore forces a low string scale and suggests that the first signatures of large extra dimensions would be stringy physics such as string resonances. We further use bounds on the string scale to place bounds on the size of the low effective Planck scale.

In Ref. \cite{Atkins:2010re}, the s-channel scattering of matter particles via the exchange of KK gravitons has been considered. It has been shown using a partial wave amplitude analysis that perturbative unitarity is typically violated at about half the effective Planck mass. The partial wave $a_J$, can be determined using ${\cal A} =16 \pi \sum_J (2J+1) a_J d^J_{\mu,\mu^\prime}$.  The different partial waves for scattering via KK gravitons in the massless limit are reproduced in Table (1). 
\begin{table*}[tbh]
\resizebox{\textwidth}{!}{
\begin{tabular}{|c|c|c|c|c|c|}  
\hline
 $\to$ & $s' s'$ & $\psi'_+\bar \psi'_- $ & $ \psi'_-\bar \psi'_+ $ & $V'_+ V'_-$ & $V'_- V'_+$ \\
\hline $s s$  & $-2/3\pi G_N s\ d^2_{0,0} $ & $ -2\pi G_N s \sqrt{1/3}\ d^2_{0,1}$
& $ -2\pi G_N s \sqrt{1/3}\ d^2_{0,-1}$  & $-4\pi G_N s \sqrt{1/3}\ d^2_{0,2} $  & $ -4\pi G_N s\sqrt{1/3}\ d^2_{0,-2} $\\
\hline $\psi_+\bar \psi_- $ &
                $ -2\pi G_N s \sqrt{1/3}\ d^2_{1,0}$ & $ -2\pi G_N s\ d^2_{1,1} $
& $  -2\pi G_N s\ d^2_{1,-1} $ &  $-4\pi G_N s\ d^2_{1,2}$ & $-4\pi G_N s \  d^2_{1,-2}$ \\
\hline $\psi_-\bar \psi_+ $ &
                $-2\pi G_N s \sqrt{1/3}\ d^2_{-1,0}$ & $ -2\pi G_N s\ d^2_{-1,1} $
& $ -2\pi G_N s\ d^2_{-1,-1} $ &   $ -4\pi G_N s\ d^2_{-1,2}$ & $ -4\pi G_N s \ d^2_{-1,-2}$ \\
\hline $V_+ V_- $ &  $ -4\pi G_N s \sqrt{1/3}\ d^2_{2,0}$ &  $ -4\pi G_N s\
d^2_{2,1}$ &$
 -4\pi G_N s\ d^2_{2,-1}  $ & $-8\pi G_N s\ d^2_{2,2}$ & $-8\pi G_N s\ d^2_{2,-2}$ \\
\hline $V_- V_+  $ &  $-4\pi G_N s\sqrt{1/3}\ d^2_{-2,0}$ &  $ -4\pi G_N s\
d^2_{-2,1}$ &$
 -4\pi G_N s\ d^2_{-2,-1}  $ & $-8\pi G_N s\ d^2_{-2,2}$ & $-8\pi G_N s\ d^2_{-2,-2}$ \\
\hline  
\end{tabular}}
\caption{Scattering amplitudes for scalars, fermions, and vector
bosons via s-channel KK graviton exchange in terms of the Wigner
$d$ functions.  $G_N=(8 \pi \bar M_P^2)^{-1}$ is Newton's constant and $s=E^2_{\rm{CM}}$ is the centre of mass energy squared. We have used the helicity basis.}  \label{t1}
\end{table*}

Each partial wave is subject to the unitarity bound  $| \mbox{Re} \ a_J|\le1/2$. Considering the $J=2$ partial wave for the scattering of a superposition of states, $|\sqrt{1/3} \sum s  s + \sum \psi_- \bar \psi_++ 2 \sum V V \rangle$, one finds
\begin{equation}\label{oneKKbound}
|a_2|=\frac{1}{320 \pi} \frac{s}{\bar M_P^2}N \le\frac12
\end{equation}
where $N=1/3 N_S + N_F + 4N_V$. Each amplitude in Table (1) can however occur via exchange of any of a very large number of KK gravitons and we will not always be in the massless limit. To take into account the exchange of a massive KK graviton with mass $m_i$, each of the amplitudes in Table (1) should be multiplied by $s/(s-m_i^2)$. Because of the very small spacing of the KK masses in models with large extra dimensions, we can approximate summing up all of the amplitudes for each $m_i$ by an integral. The number of modes with masses between $m$ and $m+dm$ is given by
\begin{equation}\label{dN}
dN= S_{n-1} m^{n-1} R^n dm 
\end{equation}
where $S_{n-1}=2 \pi^{n/2}/ \Gamma(n/2)$ is the surface of a unit-radius sphere in $n$ dimensions. Summing all the modes with masses $m_i \le \Lambda$, we find
\begin{equation}\label{sumint}
\sum_i \frac{1}{s-m_i^2} \approx \int_0^{\Lambda}\frac{1}{s-m^2}S_{n-1} m^{n-1} R^n dm .
\end{equation}
The integral clearly diverges for masses with $m_i^2=s$, and since the masses of the KK modes are so finely spaced this always occurs. This problem should be dealt with by introducing a width for the KK gravitons $\Gamma(m) \sim \frac{m^3}{\bar M_P^2}$. We can now sum up all the KK modes using
\begin{equation}\label{sumintEstar}
\int_0^{\Lambda}\frac{1}{s-m^2 + i m \Gamma(m)}S_{n-1} m^{n-1} R^n dm .
\end{equation}
In  Ref. \cite{Atkins:2010re} we cut off the integral at the energy at which unitarity is violated, i.e. we set $\Lambda=\sqrt{s}=E_\star$, since this is where the effective theory will break down. We find that with $M_D = 1$ TeV, unitarity is violated at $\sqrt{s}=E_\star = 0.46, 0.47, 0.49, 0.53, 0.56$ and $0.59$ TeV for $n=2, 3, 4, 5, 6$ and $7$ respectively. This ratio of the scales, $E_\star \sim \frac12M_D$, remains the same at higher energies, for example with $n=6$ extra dimensions with $M_D = 5$ TeV we find $E_\star = 2.8$ TeV. In other words, perturbative unitarity is violated before strong gravitational physics effects have a chance to kick in. This is one of the main results obtained in \cite{Atkins:2010re}.

The above results are robust to changes in the proportionality constant in the expression of the width $\Gamma(m) =c \frac{m^3}{\bar M_P^2}$. Our results are obtained for $c=1$, but varying $c$ by a factor of 100 in either direction only affects the bounds at the sub percent level. Finally, an expression for the decay width is given in \cite{Han:1998sg} to be
\begin{equation}\label{width}
\Gamma(\mathrm{graviton} \to \mathrm{SM}) = \frac{293}{960\pi} \frac{m^3}{\bar M_P^2}
   \end{equation}
i.e. $c \sim 0.1$ for KK gravitons with masses large enough to decay into any standard model particles. We also remark here on the suitability of the Breit-Wigner width. The spacing between KK modes is approximately $\Delta m \simeq 1/R = M_D (M_D/\bar M_P)^{(2/n)}$, and so for $n \ge 2$ and KK masses at the TeV scale we have $\Gamma \ll \Delta m$. There is no overlap between the resonant modes and since also $\Delta m \ll m$ the use of the Breit-Wigner width is valid.

There are essentially two forms of divergent behaviour in the sum over KK modes in Eq. (\ref{sumint}) and it is important to understand which is contributing to the unitarity problem above. Firstly there are divergences appearing from the pole region where $s=m^2$, and secondly there are divergences coming from the infinite sum of the high density of states for $n>1$.  The first is an IR divergence and the second is a UV divergence. We may tame the IR divergence by introducing a width, as above, however the UV divergence can only be dealt with by introducing some form of cutoff to the number of KK modes, to which the physics is highly sensitive. A common method used to find the cutoff in such an effective theory is to calculate the lowest energy at which unitarity is violated and associate this with the cutoff, as we have done above. However, in this instance, it is rather hard to separate the effects of these two divergences in such a calculation. The problem lies in the fact that when we cut off the integral (\ref{sumintEstar}) at $\Lambda=\sqrt{s}$, we do not include modes above the pole which normally appear with opposite sign and begin to lower the amplitude. Hence when we strictly cut off the integral at the point at which unitarity is violated, we see mostly the resonant behaviour near the pole, and it is therefore hard to determine in this way unitarity problems stemming from the UV divergence. The calculation is also therefore extremely sensitive to the cutoff.

It is actually quite clear to see that the unitarity problem is coming from the resonant behaviour of on shell virtual gravitons since it also shows up at a low energy for $n=2$. In this case, the UV behaviour of the KK sum is only logarithmically divergent and so one would expect that any unitarity problems stemming from the UV divergence of the KK sum would appear at much higher energies.

The use of the width in the propagator introduces an imaginary part to the amplitude and we can also apply the unitarity bound to this, i.e. $| \mbox{Im} \ a_J| <1$. It is clear that a bound derived from the imaginary part is only produced by the on shell behaviour of the KK gravitons and does not encapsulate any of the UV divergence. It is therefore far less sensitive to the cutoff. Applying the bound in this way, we find that unitarity is violated at  $\sqrt{s}=E_\star = 1.02, 0.89, 0.84, 0.83, 0.83$ and $0.84$ TeV for $n=2, 3, 4, 5, 6$ and $7$ respectively. Again these bounds are not particularly sensitive to the proportionality constant used in the expression for $\Gamma[m]$. Also the bounds scale with $M_D$, so in general we find that unitarity is violated at $E_\star \sim 0.8M_D$. We emphasise again that all these bounds are stemming from resonant behaviour as on shell gravitons are exchanged  and not a result of the UV divergence from the sum over KK modes. However the bound on the imaginary part is far less sensitive to the cutoff and therefore seems a more appropriate tool to use to find the scale of unitarity violation.

Another way to deal with the on shell part of the amplitude is via the $i \epsilon$ prescription and taking the principle value of the integral, as is done in \cite{Han:1998sg}. Here the integral (\ref{sumint}) is done analytically using dimensional regularisation and an imaginary part is found proportional to $\pi s^{(n-2)/2}$. Bounding this for individual particle in and out states one finds $E_\star \sim 1.5 M_D$ which can be reduced further by again considering scattering of the state $|\sqrt{1/3} \sum s  s + \sum \psi_- \bar \psi_++ 2 \sum V V \rangle$ to $E_\star \sim 0.15 M_D$. However since the KK gravitons are not stable the true way to deal with this is to introduce a width as above and we therefore maintain that the more conservative bound of $E_\star \sim 0.8 M_D$ is the most relevant.

To summarise, if we insist that we should strictly cut off the sum over KK modes at the energy at which unitarity is violated, we find by bounding the real part of the amplitude that unitarity breaks down at around $\frac12 M_D$. This bound appears due to the resonance behaviour of on shell KK gravitons being exchanged and is extremely sensitive to the point at which we cut off the modes. We therefore prefer to bound the imaginary part of the amplitude which is fairly insensitive to the cutoff and we find in general that unitarity breaks down at approximately $0.8 M_D$. 

One might expect that strong gravitational effects appear at this scale and fix the unitarity problem via higher orders in perturbation theory. However, it is shown in  \cite{Han:1998sg, Giudice:2003tu} that by naive dimensional analysis, one expects gravity to become strong at around
\begin{equation}\label{gravstrong}
\Lambda_{\mbox{strong}}=\left[ \Gamma\left(2+n/2\right)\right]^{1/(2+n)}(4\pi)^{\frac{4+n}{4+2n}} M_D.
\end{equation}
For any number of extra dimensions, it is found that $\Lambda_{\mbox{strong}}>7.2M_D$. Therefore it is very unlikely that higher orders in perturbation theory will be able to fix the breakdown of unitarity. 

This problem can easily be solved if there is new physics around $E_\star$ (or below). An obvious candidate is string theory and we identify the string scale $M_S$ with $E_\star$. The fact that the breakdown of unitarity forces the string scale to appear below the quantum gravity scale fits nicely with the relationship between these two quantities derived in specific string models. In Ref. \cite{Antoniadis:1998ig} the idea that string theory provides a natural framework for models with large extra dimensions was first proposed. Here it was shown that  the only perturbative string theory with weak scale string tension, realising low scale gravity, must be a type I theory of open and closed strings with an $\mathcal{O}(1)$ string coupling. With $n$ large extra dimensions, the remaining $6-n$ dimensions must be of the string size. Given such a string realisation with the Standard Model embedded on a D-brane configuration, it was further shown that a relation exists between the effective Planck scale and the string scale, $M_S \sim \alpha^{2/(n+2)}M_D$, where $\alpha=g_S/ 4\pi$ is the string coupling. The spacing between these two scales is therefore dependent on the string coupling and $n$. It is shown for a toy model in \cite{Cullen:2000ef}, that for $n=6$ the relation is given by
\begin{equation}\label{peskin}
\frac{M_D}{M_S} = \left( \frac{1}{\pi} \right)^{1/8} \alpha^{-1/4}
\end{equation}
and with the choice $g_s=1/2$ we find that $M_D/M_S=1.9$. For the extreme choice $\alpha =1/137$, $M_D/M_S=3.0$. This relation is model dependent however we have shown from unitarity arguments that $M_S$ is required to be at $0.8 M_D$ or lower. From this we can infer that the dominant first signatures of large extra dimensions would be stringy physics such as string resonances \cite{Cullen:2000ef, Anchordoqui:2008di}, before we see strong gravitational effects.~\footnote{Besides string resonances, there are often other predictions at lower energies, such as TeV KK excitations of Standard Model gauge bosons~\cite{Antoniadis:1990ew}, or anomalous $U(1)$'s~\cite{Anchordoqui:2011eg}, that could be the first manifestations of low scale strings.}

Given the requirement $M_S \lesssim 0.8 M_D$ we can use bounds on the string scale to bound the effective Planck scale. We therefore briefly review the most current bounds on $M_S$. String effects at energies below the string scale can be characterised by effective operators for contact interactions \cite{Cullen:2000ef, Antoniadis:2000jv}. For example in models where matter fermions sit on the same set of branes these would take the form of dimension-8 operators such as
\begin{equation}\label{dim8}
\frac{g_s}{M_S^4}(\bar\psi \partial \psi)^2
\end{equation}
and for fermions living on brane intersections the dimension-6 operator
\begin{equation}\label{dim6}
\frac{g_s}{M_S^2}(\bar\psi  \psi)^2.
\end{equation}
Model dependent bounds derived from the dimension eight operators are rather weak \cite{Bourilkov:2000ap, Antoniadis:2000jv} and we will not consider them further here. In Ref. \cite{Antoniadis:2000jv} it is shown that the dimension six operators can be generically parameterised as
\begin{equation}\label{dim6param}
\mathcal{L}_{eff} = \frac{4 \pi}{(1+\epsilon)\Lambda}\sum_{a,b=L,R} \eta_{ab} \bar\psi_a \gamma^\mu \psi_a \bar\psi_b' \gamma_\mu \psi_b'
\end{equation}
with $\epsilon=1$ $(0)$ for $\psi = \psi'$ $(\psi \neq \psi')$. Then for the specific model considered in \cite{Antoniadis:2000jv} one can identify
\begin{equation}\label{Lambda}
\Lambda \simeq \sqrt{\frac{4 \pi}{0.59 g_s}}M_S
\end{equation}
which is the quantity normally identified as $\Lambda_{VV}^+$. The most stringent bound on this quantity comes from LEP: $\Lambda_{VV}^+ > 21.7$ TeV \cite{Alcaraz:2006mx}. Choosing $g_S \simeq 1/2$ we find $M_S > 3.3$ TeV. With the extreme choice of $g_S/4\pi =1/128$ we find $M_S > 1.4$ TeV. Or, using the results from \cite{Barbieri:1999tm} there is less ambiguity over the choice of $g_S$ and we find with $g_S \simeq 1/2$, $M_S > 3.1$ TeV and with $g_S = 0.425$, $M_S > 2.2$ TeV.

The most general model independent bounds can be placed on the string scale by searching for the production of dijet resonances at hadron colliders \cite{Cullen:2000ef, Anchordoqui:2008di}. The current tightest bound comes from recent data from the CMS experiment operating at a centre of mass energy $\sqrt{s}=7$ TeV and with $1 \: \mathrm{fb}^{-1}$ of data, producing the model independent bound $M_S > 4.00$ TeV \cite{Collaboration:2011ns}. The ATLAS experiment has not yet produced direct bounds on $M_S$ but in \cite{Aad:2011aj} they produce a lower bound of $2.10$ TeV on the axigluon mass which can be roughly translated into the same limit on $M_S$ \cite{Cullen:2000ef}. 

In order to fix the unitarity problem we expect string physics to appear at the scale of unitarity violation, i.e. $M_S \simlt E_\star \sim 0.8 M_D.$ We can therefore use bounds on $M_S$ to place bounds on the effective Planck scale. If we take the model independent bound $M_S>4.00$ TeV we find
\begin{equation}\label{MD1}
M_D\gtrsim 5 \: \mathrm{TeV}.
\end{equation}
Choosing the specific model dependent bound $M_S > 3.3$ TeV, coming from dimension-6 operators, we find
\begin{equation}\label{MD2}
M_D\gtrsim 4.1 \: \mathrm{TeV}.
\end{equation}

Previously such a relationship between the string scale and the effective Planck scale was not thought to be necessary but was just a result of specific string models. We have found that the requirement that string physics must appear at the scale of unitarity violation forces such a relationship and therefore bounds the quantum gravity scale Eq.(\ref{MD1}). This pushes further up, in particular, the relevant scale of micro black hole production, at energies beyond the LHC reach independently of other theoretical arguments. 

With the string scale appearing below the effective Planck scale, the dominant collider signatures of large extra dimensions will be stringy physics such as production of string resonances. With large amounts of data constantly being collected at the LHC, if large extra dimensions do exist in nature so that the string scale is not very far from the electroweak scale, we will surely see evidence of them soon and be able to experimentally probe the exciting world of string theory.

\section*{Acknowledgements}
The work of I.A. was supported in part by the European Commission under the ERC Advanced
Grant 226371 and the contract PITN-GA-2009-237920. The work of M.A. was supported by the Science and Technology Facilities Council [grant number ST/1506029/1]. The work of X.C. is supported in part by the European Cooperation in Science and Technology ({\sc COST}) action {\sc MP0905} ``Black Holes in a Violent Universe.''


\bigskip



\baselineskip=1.6pt
\begin{thebibliography}{99}

\bibitem{ArkaniHamed:1998rs}
  N.~Arkani-Hamed, S.~Dimopoulos and G.~R.~Dvali,
  Phys.\ Lett.\  B {\bf 429}, 263 (1998)
  [arXiv:hep-ph/9803315];

\bibitem{Antoniadis:1998ig}
  I.~Antoniadis, N.~Arkani-Hamed, S.~Dimopoulos and G.~R.~Dvali,
  Phys.\ Lett.\  B {\bf 436}, 257 (1998)
  [arXiv:hep-ph/9804398].

\bibitem{Atkins:2010re}
  M.~Atkins and X.~Calmet,
  Eur.\ Phys.\ J.\  C {\bf 70}, 381 (2010)
  [arXiv:1005.1075 [hep-ph]];
  Phys.\ Lett.\  B {\bf 695}, 298 (2011)
  [arXiv:1002.0003 [hep-th]].

\bibitem{Han:1998sg}
  T.~Han, J.~D.~Lykken and R.~J.~Zhang,
  Phys.\ Rev.\  D {\bf 59}, 105006 (1999)
  [arXiv:hep-ph/9811350];
  G.~F.~Giudice, R.~Rattazzi and J.~D.~Wells,
  Nucl.\ Phys.\  B {\bf 544}, 3 (1999)
  [arXiv:hep-ph/9811291].


\bibitem{Giudice:2003tu}
  G.~F.~Giudice, A.~Strumia,
  Nucl.\ Phys.\  {\bf B663}, 377-393 (2003).
  [hep-ph/0301232].

\bibitem{Anchordoqui:2008di}
  L.~A.~Anchordoqui, H.~Goldberg, D.~Lust, S.~Nawata, S.~Stieberger, T.~R.~Taylor,
  Phys.\ Rev.\ Lett.\  {\bf 101}, 241803 (2008).
  [arXiv:0808.0497 [hep-ph]].


\bibitem{Cullen:2000ef}
  S.~Cullen, M.~Perelstein, M.~E.~Peskin,
  Phys.\ Rev.\  {\bf D62}, 055012 (2000).
  [hep-ph/0001166].


\bibitem{Antoniadis:1990ew}
  I.~Antoniadis,
  Phys.\ Lett.\  B {\bf 246} (1990) 377.


\bibitem{Anchordoqui:2011eg} See for instance:
  L.~A.~Anchordoqui, I.~Antoniadis, H.~Goldberg, X.~Huang, D.~Lust and T.~R.~Taylor,
  arXiv:1107.4309 [hep-ph]; and references therein.


\bibitem{Antoniadis:2000jv}
  I.~Antoniadis, K.~Benakli, A.~Laugier,
  JHEP {\bf 0105}, 044 (2001).
  [hep-th/0011281].


\bibitem{Bourilkov:2000ap}
  D.~Bourilkov,
  Phys.\ Rev.\  {\bf D62}, 076005 (2000)
  [hep-ph/0002172].

\bibitem{Alcaraz:2006mx}
  J.~Alcaraz {\it et al.}  [ALEPH Collaboration and DELPHI Collaboration and
                  L3 Collaboration and LEP Electroweak Working Group],
  arXiv:hep-ex/0612034.

\bibitem{Barbieri:1999tm}
  R.~Barbieri, A.~Strumia,
  Phys.\ Lett.\  {\bf B462}, 144-149 (1999).
  [hep-ph/9905281].


\bibitem{Collaboration:2011ns}
 S.~Chatrchyan {\it et al.}  [CMS Collaboration],
  [arXiv:1107.4771 [hep-ex]].


\bibitem{Aad:2011aj}
  G.~Aad {\it et al.} [ATLAS Collaboration],
  New J.\ Phys.\  {\bf 13}, 053044 (2011).
  [arXiv:1103.3864 [hep-ex]].


\end{thebibliography}
\end{document}